\def\be{\begin{equation}}
\def\ee{\end{equation}}
\def\ba{\begin{eqnarray}}
\def\nn{\nonumber}
\def\ea{\end{eqnarray}}

\def\a{\alpha}
\def\b{\beta}

\def\d{\delta}

\def\f{\phi}

\def\m{\mu}
\def\n{\nu}
\def\o{\omega}
\def\p{\pi}
\def\q{\theta}

\def\s{\sigma}
\def\t{\tau}

\def\G{\Gamma}

\def\O{\Omega}

\def\Ft{\tilde{F}}
\def\in{\infty}
\def\nt{\tilde{\nu}}
\def\mt{\tilde{\mu}}

\def\inti{\int_{-\infty}^{t_i^{-1}(T)}}
\def\intij{\int_{-\infty}^{t_j^{-1}(t_i(\tau_i))}}

\def\ga{\gamma }
\def\intin{\int_{-\infty }^{\infty }}
\def\intiz{\int_{0}^{\infty }}

\def\cw{\chi_{\omega}}
\def\ck{\chi_{k}}
\def\cwp{\chi_{\omega '}}
\def\cwo{\chi_{\omega}^{(1)}}
\def\cwop{\chi_{\omega '}^{(1)}}
\def\cwt{\chi_{\omega}^{(2)}}
\def\cwtp{\chi_{\omega '}^{(2)}}

\def\Qt{\tilde{Q}}
\def\Qot{\tilde{Q}_1}
\def\Qtt{\tilde{Q}_2}
\def\etat{\tilde{\eta}}
\def\etaot{\tilde{\eta}_1}
\def\etatt{\tilde{\eta}_2}

\def\ika{\frac{ik}{a}}
\def\pka{\frac{\pi k}{a}}
\def\pkta{\frac{\pi k}{2a}}

\def\gt{\tilde{\gamma}}

\newskip\humongous \humongous=0pt plus 1000pt minus 1000pt

\newif\ifdtup

\def\a{\alpha}

\def\ha{{1\over 2}}

\def\(#1){(\ref{#1})}
\newcommand{\bb}{\bibitem}

\documentstyle[12pt]{article}

\makeatletter                    %allow @ in command names
\@addtoreset{equation}{section}  %reset equation count in each section
\makeatother                     %disallow @ in command names

\begin{document}
\title{Stochastic Theory of Accelerated Detectors in a Quantum Field}
\author{Alpan Raval\thanks{e-mail: raval@umdhep.umd.edu},  B. L. Hu\thanks{
e-mail: hu@umdhep.umd.edu}\\
{\small Department of Physics, University of Maryland,
 College Park, Maryland 20742}\\
{\small School of Natural Sciences, Institute for Advanced Study, Princeton,
NJ 08540, USA}\\
{\small Department of Physics, Hong Kong University of Science and
Technology,} \\
{\small Clear Water Bay, Kowloon, Hong Kong}\\
 James Anglin\thanks{e-mail: anglin@tdo-serv.lanl.gov}\\
 {\small  Theoretical Astrophysics, MS B-288,
Los Alamos National Laboratory, NM 87545}}
\date{\small {\it (IASSNS-HEP-95/39, UMDPP 95-131, LA-UR-95-1736,
Submitted to Phys. Rev. D on May 23, 1995)}}
\maketitle

\begin{abstract}
We analyze the  statistical mechanical properties of n-detectors in arbitrary
states of motion interacting with each other via a quantum field.
We use the open system concept and the influence functional method to
calculate the influence of quantum fields on detectors in motion,
and the mutual influence of detectors via fields.
We discuss the difference between self and mutual impedance,
advanced and retarded noise,  and the relation of noise-correlations and
dissipation-propagation.
The mutual effects of detectors on each other can be studied from
the Langevin equations derived from the influence functional, as it
contains the backreaction of the field on the system self-consistently.
We show the existence of general fluctuation-dissipation relations, and
for trajectories without event horizons, correlation-propagation relations,
which succinctly encapsulate these quantum statistical phenomena.
These findings serve to clarify some existing confusions on the accelerated
detector problem. The general methodology presented here could also serve
as a platform to explore the quantum statistical properties of particles
and fields, with practical applications in atomic and optical physics problems.

\end{abstract}

\newpage

\section{Introduction}

The physics of accelerated detectors became an interesting subject of
investigation when Unruh showed that a detector moving with uniform
acceleration sees the vacuum state of some quantum field in Minkowski
space as a thermal bath with temperature $T_U = \hbar a /2 \pi c k_B$
\cite{Unr76}. This seminal work which uses the structure of quantum
field theory of Rindler space explored by Fulling \cite{Fulling},
and the Bogolubov transformation ideas invented by Parker for cosmological
particle creation \cite{Par69} earlier,  draws a clear parallel with the
fundamental discovery of Hawking radiation in black holes \cite{Hawking75}.
The discovery of Unruh effect (see also Davies \cite{Davies})
sets off the first wave of activities on this subject.
The state-of-the-art understanding of the physics of this problem
in this first stage of work is represented by the paper of Unruh and Wald
\cite{UnrWal}. We refer the readers to the reviews of Sciama, Candelas and
Deutsch \cite{SCD}, Tagaki \cite{Tak} and Ginzburg and Frolov \cite{GinFro}.

The second stage of investigation on this problem was initiated by the
inquiry of Grove \cite{Grove}, who challenged the prevailing view and
asked the question whether the detector actually radiates. This was answered
in the negative by an inspiring paper of Raine, Sciama and Grove \cite{RSG}
(RSG henceforth) who considered an exactly solvable harmonic oscillator
detector
model and analyzed what an inertial observer sees
in the forward light cone of the accelerating detector via a Langevin equation.
Unruh \cite{Unr92} performed an independent
calculation and concurred with the findings of RSG to the extent that the
energy-momentum tensor of the field as modified by the presence of the
accelerating detector vanishes over most of the spacetime (except on the
 horizons). However, he also showed the existence of extra terms in the
two-point function of the field beyond its value in the absence of the
accelerating detector, and argued that these terms would contribute to
the excitation of a detector placed in the forward light cone.
These terms were missed out in RSG. Following
these exchanges, there was a recent renewed interest in this problem,
notably the series of papers by Massar, Parantani and Brout \cite{MPB} (MPB),
who gave a detailed analysis via Hamiltonian quantum mechanics
of the two-point function and pointed out that the missing terms contribute
to a polarization cloud around the accelerating detector;
Hinterleitner \cite{Hin}, who independently discussed the backreaction of the
detector on the field using a slightly different yet exactly solvable model and
arrived at similar conclusions to MPB; and Audretsch and M\"uller %and Holzmann
\cite{AM}, who explored nonlocal pair correlations in accelerated detectors.
%and the relation between energy shifts and relaxation
%rates for a two-level atom in a reservoir. %from the point of view of the
%%balance between radiation reaction and vacuum fluctuations.
However, the physical significance of the polarization cloud,
its connection to the noise experienced by another detector, and to
the inherent correlations in the free Minkowski vacuum, remain largely
unexplored.

Beginning with this work we would like to add a new dimension to this problem
and open up the third stage of investigation. The new emphasis is in exploring
the statistical mechanics of particles and fields, and in particular, moving
detectors on arbitrary trajectories. We analyze the stochastic properties of
quantum fields and discuss this problem in terms of quantum noise,
correlation and dissipation. We use the open
system concept and the influence functional method to treat a system of
n detectors interacting with a scalar field. This method enables one to examine
the influence of detectors in motion on quantum fields, the mutual
influences of detectors via fields, as well as the backreaction of fields
on detectors in a self-consistent manner.

As explained earlier \cite{Banff}, the influence functional method is a
generalization of the powerful effective action method in quantum field theory
for treating backreaction problems, which also incorporates statistical
mechanics notions such as noise, fluctuations, decoherence and dissipation.
Indeed, one of us has long held the viewpoint that \cite{HuPhysica,HuEdmonton},
to get a more profound
understanding of the meaning of Unruh and Hawking effects and the black hole
information and backreaction problems one cannot be satisfied with the
equilibrium thermodynamics description. It is necessary to probe deeper
into the statistical properties of quantum fields, their correlations and
dynamics, coherence and decoherence of the particle-field system,
the relation of quantum noise and thermal radiance, fluctuation-dissipation
relations, etc.  Earlier investigation of correlation and dissipation
in the Boltzmann-BBGKY scheme \cite{CH88,dch,cddn}
and the properties of noise and fluctuations in the Langevin
framework \cite{HPZ,HM2,nfsg} are essential preparations for tackling
such problems at a deeper level.\footnote{Our view is thus most akin to
that espoused by Sciama \cite{Sciama}.}

With this theoretical perspective in mind, we have recently begun a systematic
study of the accelerated detector problem \cite{Anglin,HM2}. We show that
thermal radiance can be
understood as originating from quantum noise under different kinematical
(moving detector) and dynamical (cosmology) excitations. The aim of this paper
is to  1) show on both the conceptual and technical levels the power and
versatility of this new method, 2) settle some open questions and clarify
some existing confusions, such as the existence of radiation and polarization,
solely from an analysis of detector response, 3) introduce new concepts
such as self and mutual impedance, advanced and retarded noise,
fluctuation-dissipation and correlation-propagation relations using the
accelerated detectors problem as example,
%into the existing framework of detector response via kinematical excitation,
and finally 4) suggest new avenues of investigations into
the statistical mechanics of particle and fields, including black hole physics.

Employing a set of coupled stochastic equations for the detector dynamics,
we analyze the influence of an accelerated detector on a probe which
is not allowed to
causally influence the accelerated detector itself. We find, as did
\cite{RSG,Unr92,MPB} that most of the terms in the correlations of the
stochastic force acting on the probe cancel each other. This cancellation
is understood in the light of a correlation-propagation relation, derived as
a simple construction from the fluctuation-dissipation relation
for the accelerated detector. Such a relation can be equivalently viewed
as a construction of the free field two-point function for each point on
either trajectory from the two-point function along the uniformly accelerated
trajectory alone. The remaining terms, which contribute to
the excitation of the probe, are shown to represent correlations
of the free field across the future horizon of the accelerating
detector. In this problem, the dissipative properties of either detector
remain unchanged by the presence of the other. This happens because the
probe cannot influence the accelerated detector. However, the stochastic
 force acting on the probe plays a non-trivial
role.

We also consider the problem of two inertial detectors
which can backreact on each other. This mutual backreaction changes the
self-impedance functions of these detectors, and introduces mutual
impedances as well. The dissipative properties of each detector are thus
altered due to the presence of the other one. This physical effect is in a
sense complementary to the effects manifested in the accelerated detector
problem, where the probe does not backreact on the accelerated detector.

The paper is organized as follows. In Section 2 we develop the influence
functional formalism describing the influence of a massless scalar field on
a system of an arbitrary number of detectors moving on arbitrary trajectories.
The field modes are integrated out in this formalism, and effective stochastic
equations of motion for the various detectors are obtained. In Section 3 we
consider some applications of this formalism to three simple cases, the primary
one being the analysis of two inertial detectors coupled to the same
quantum field. In Section 4 we treat the RSG excitation of a probe in the
presence of a
uniformly accelerating detector. In Section 5 we show the existence of
fluctuation-dissipation relations governing the detector system. These
relations are used as a starting point for obtaining more general relations
between the correlations of various detectors and the radiation mediated by
them. Such relations are also discussed in the specific context of the RSG
model. Finally, in the Appendix, we point out problems associated with
the uncorrelated detector-field initial state in a minimally coupled model,
and argue that these problems are removed in a derivative coupling model.
We present a simple prescription for switching from one model to the other.

\section{Scalar Electrodynamics or Minimal Coupling Model}

The paper by Raine, Sciama and Grove uses the scalar
electrodynamic or ``minimal'' coupling of oscillators to a scalar field
in 1+1 dimensions.  This coupling provides a positive definite
Hamiltonian, and is of interest because it resembles the actual
coupling of charged particles to an electromagnetic field. In this
section, we derive the influence functional describing the effect of
a scalar field on the dynamics of an arbitrary number of detectors
modelled as minimally coupled oscillators. The detectors
move along arbitrary trajectories. We assume that the field and the system of
detectors are initially decoupled from each other, and that the field is
initially in the Minkowski vacuum state. The formalism can be simply extended
to
higher dimensions, and to different choices of initial state for the field.
 We also obtain coupled Langevin equations for the detector system.

\subsection{Influence functional for N arbitrarily moving detectors}

Consider $N$ detectors $i=1,..N$ in $1+1$ dimensions with internal oscillator
coordinates $ Q_i(\tau_i), $ and trajectories $(x_i(\tau_i), t_i(\tau_i))$
, $\tau_i$ being a parameter along the trajectory of detector $i$.
In the following analysis, we do not need to assume that $\t _i$ is the proper
time, although this is, in most cases, a convenient choice. However, we will
assume hereafter that the trajectories $(t_i(\tau_i),x_i(\tau_i))$
are smooth and that the parameters $\tau_i$ are chosen such that
 $t_i(\tau_i)$ is a strictly increasing function of $\tau_{i}$.

The detectors are coupled to a massless scalar field $\phi(x,t)$ via the
interaction action
\be
S_{int} = \sum _i e_i \int _{-\infty }^{t_i^{-1}(T)} d\tau_i   s_{i}(\tau_{i})
\frac{d Q_i}{d\tau_i} \phi (x_i(\tau_i), t_i(\tau_i)).
\ee
Here, $T$ is a global Minkowski time coordinate which defines a spacelike
hypersurface, $e_i$ denotes the coupling
constant of detector $i$ to the field, $s_{i}(\tau_{i})$ is the switching
function for detector $i$ (typically a step function), and $t_{i}^{-1}$ is the
inverse function of $t_{i}$. $t_i^{-1}(T)$ is therefore the value of $\t _i$ at
the point of intersection of the spacelike hypersurface defined by $T$ with the
trajectory of detector $i$. Note that the strictly increasing property of
$t_i(\t _i)$ implies that the inverse, if it exists, is unique.

The action of the system of detectors is
\be
S_{osc} = \ha \sum _i \int _{-\infty }^{t_i^{-1}(T)} d\tau_i
 [(\partial_{\tau_i}Q_{i})^2 - \O _{i}^2 Q_{i}^2].
\ee
The scalar field action is given by
\be
S_{field} = \frac{1}{2} \int_{-\infty }^{T} dt \int dx  [(\partial_t\f )^2 -
(\partial_x\f )^2]
\ee
and the complete action
\be
S = S_{field} + S_{osc} + S_{int} .
\ee

Expanding the field in normal modes,
\be
\phi(x,t) = \sqrt {\frac{2}{L}} \sum '_k [ q_k^+ (t) \cos  kx + q_k^- (t) \sin
kx]
\ee
where $\sum '_k$ denotes that the summation is restricted to the upper half
$k$ space, $k>0$.
Then the action for the scalar field is given by ($\sigma = +,-$)
\be
S_{field} = \frac{1}{2} \sum '_{k,\sigma} [ (\dot q_k^\sigma)^2 - \omega_k^2
q_k^2]
\ee
and the interaction action is
\ba
S_{int} &=& \sum _{i}e_i \sqrt {\frac{2}{L}} \int_{-\infty }^{t_i^{-1}(T)}
d\tau_i
\frac{dQ_i}{d\tau_i} \times     \nn \\
& &\sum '_{k} [ q_k^+ (t_i(\tau_i)) \cos  kx_i(\tau_i)
+ q_k^- (t_i(\tau_i)) \sin  kx_i(\tau_i) ] s_i(\tau_i)    \nn \\
       &=& \sum _{i}e_i \sqrt {\frac{2}{L}} \int_{-\infty }^{\infty }dt
        \int_{-\infty }^{t_i^{-1}(T)} d\tau_i \d (t-t_i(\tau_i))
        \frac{dQ_i}{d\tau_i} \times  \nn \\
        & &\sum '_{k} [ q_k^+ (t) \cos  kx_i(\tau_i) + q_k^- (t) \sin
kx_i(\tau_i) ]
        s_i(\tau_i).
\ea
We have $t_{i}(\tau_i) < T$, which follows from $\tau_i < t_i^{-1}(T)$ and
the property that $t_i(\tau_i)$ is a strictly increasing function. Hence
we may replace the upper limit of the $dt$ integration by $T$. This
replacement leads to the expression:
\be
S_{int} = - \sum '_{k,\sigma} \int_{-\infty }^{T} dt J_k^\sigma (t)
q_k^\sigma (t)
\ee
where
\be
J_k^{\sigma}(t) = -\sum_i e_i \sqrt{\frac{2}{L}} \int_{-\in }^{t_i^{-1}(T)}
d\tau_i
\delta(t-t_i(\tau_i)) \frac{dQ_i}{d\tau_i} u_{k}^{\sigma}(\tau_i) s_i(\tau_i)
\ee
and
\be
u_{k}^+(\tau_i) =  \cos  kx_i(\tau_i) ;~~~u_{k}^-(\tau_i) =  \sin
kx_i(\tau_i).
\ee

The action $S_{field} + S_{int}$ therefore describes a system of decoupled
harmonic oscillators each driven by separate source terms. The zero
temperature influence functional (corresponding to the initial state of the
field being the Minkowski vacuum) for this system has the form \cite{HPZ}:
\be
{\cal F}[J,J'] = \exp \{- {1\over \hbar} \sum '_{k,\sigma} \int_{-\in }^T ds
\int_{-\in }^s ds'[J_k^{\sigma}(s)-J_k^{'\sigma}(s)] [\zeta_k (s,s')
J_k^{\sigma}(s')
 -\zeta_k^*(s,s')J_k^{'\sigma} (s')] \}
\ee
where
\be
\zeta_k \equiv \nu_k + i\mu_k = \frac{1}{2\omega_k}e^{-i\omega_k(s-s')}.
\ee
If the field is initially in a thermal state, the influence functional has
the same form as above, and the quantity $\zeta_k $ becomes
\be
\zeta_k = \frac{1}{2\omega_k}[\coth(\frac{\beta \o
_k\hbar}{2})\cos\omega_k(s-s')
-i\sin\omega_k(s-s')],
\ee
$\beta$ being the inverse temperature.
We shall restrict our attention to the zero temperature case.

Substituting for the $J_k^{\sigma}$'s in the influence functional, and carrying
out the $\delta$-function integrations, one obtains

\ba
{\cal F}[\{Q\};\{Q'\}] &=& \exp -{1\over \hbar}\{ \sum_{i,j=1}^{N} \int_{-\in
}^
{t_i^{-1}(T)}d\tau_i s_i(\tau_i) \int_{-\in }^{t_j^{-1}(t_i(\tau_i))} d\tau'_j
s_j(\tau'_j) [\frac{dQ_i }{d\tau_i}-\frac{dQ'_i}{d\tau_i}] \times \nn \\
& &[ Z_{ij}  (\tau_i, \tau '_j) \frac{dQ_j}{d\tau '_j}
- Z_{ij}^*(\tau_i, \tau '_j) \frac{dQ'_j}{d\tau '_j} ]\}
\ea
where
\be
 Z_{ij}(\tau_i, \tau '_j) =\frac{2}{L} e_i e_j \sum '_{k,\sigma}
\zeta_k(t_i(\tau_i),t_j(\tau'_j)) u_{k}^\sigma (\tau_i) u_{k}^\sigma (\tau
'_j).
\ee
In the above, the continuum limit in the mode sum is recovered through
the replacement $\sum '_k \rightarrow \frac{L}{2\p } \int_0^{\in }dk$. We then
obtain, after substituting for $u_k^{\sigma }$ and $\zeta_k$,
\be
 Z_{ij}(\tau_i, \tau '_j) =\frac{e_i e_j}{2\p } \int_0^{\infty } \frac{dk}{k}
e^{-ik(t_i(\tau_i)-t_j(\tau'_j))} \cos k(x_i(\tau_i) - x_j(\tau'_j)).
\ee
In this form, $Z_{ij}$ is proportional to the two point function of the
free scalar field in the Minkowski vacuum, evaluated for the two points
lying on trajectories $i$ and $j$ of the detector system. It obeys the
symmetry relation
\be
 Z_{ij}(\tau_i, \tau '_j) =  Z_{ji}^{\ast }(\tau'_j, \tau _i)
\ee

Corresponding to ($2.12$), we may also split $Z_{ij}$ into its real and
imaginary parts. Thus we define
\be
Z_{ij}(\t _i,\t '_j) = \tilde{\n }_{ij}(\t _i,\t '_j) + i\tilde{\m }_{ij}(\t
_i,\t '_j)
\ee
where
\ba
\nt _{ij}(\t _i,\t '_j) &=& \frac{e_i e_j}{2\p } \int_0^{\in } \frac{dk}{k}
\cos k(t_i(\tau_i)-t_j(\tau'_j)) \cos k(x_i(\tau_i) - x_j(\tau'_j))   \nn \\
\mt _{ij}(\t _i,\t '_j) &=& -\frac{e_i e_j}{2\p } \int_0^{\in } \frac{dk}{k}
\sin k(t_i(\tau_i)-t_j(\tau'_j)) \cos k(x_i(\tau_i) - x_j(\tau'_j)).
\ea
$\nt $ and $\mt $ are proportional to the anticommutator and the
commutator of the field in the Minkowski vacuum, respectively.

The quantities $Z_{ij}$ are also conveniently expressed in terms of
advanced and retarded null coordinates $v_i(\t _i) = t_i(\t _i) + x_i(\t _i)$
and $u_i(\t _i) = t_i(\t _i) - x_i(\t _i)$, as
\be
Z_{ij}(\t _i,\t '_j) = Z_{ij}^{a}(\t _i,\t '_j) + Z_{ij}^{r}(\t _i,\t '_j)
\ee
where
\ba
Z_{ij}^{a}(\t _i,\t '_j) &=& \frac{e_ie_j}{4\p }\intiz \frac{dk}{k} e^{-ik(v_i
(\t _i)-v_j(\t '_j))}  \nn \\
Z_{ij}^{r}(\t _i,\t '_j) &=& \frac{e_ie_j}{4\p }\intiz \frac{dk}{k} e^{-ik(u_i
(\t _i)-u_j(\t '_j))}
\ea
and the superscripts $a$ and $r$ denote advanced and retarded
respectively.\footnote{
The terminology `advanced' and `retarded' refers to the null coordinates.
Equivalently, they can be called `left-moving' and `right-moving',
respectively,
when the sense of motion refers to the future direction in time. This
terminology
is used in wave theory and string theory.}
Similar decompositions for $\nt _{ij}$ and $\mt _{ij}$ thus follow.

The influence functional, along with the action for the detector system,
can be employed to obtain the propogator for the density matrix of the
system of detectors. This propogator will contain complete information
about the dynamics of the detectors. However, we shall take the alternative
approach of deriving Langevin equations for the detector system in order
to describe its dynamics.

\subsection{Langevin equations}

In this subsection, we wish to derive the effective stochastic equations of
motion for the $N$-detector system. In the previous subsection, we
integrated out the field degrees of freedom. The effect of this is to
introduce long-range interactions between the various detectors.

Going back to the form ($2.11$) for the influence functional, we define
the centre of mass and relative variables
\ba
J_k^{+\s }(s) &=& (J_k^{\s }(s) + J_k^{'\s }(s))/2   \nn \\
J_k^{-\s }(s) &=& J_k^{\s }(s) - J_k^{'\s }(s).
\ea
Correspondingly, we also find it convenient to define
\ba
Q^+_{i}(\t _i) &=& (Q_i(\t _i) + Q' _i(\t _i))/2   \nn \\
Q^-_{i}(\t _i) &=& Q_i(\t _i) - Q' _i(\t _i).
\ea
Then Equation ($2.11$) yields
\ba
\mid {\cal F}[J,J']\mid  &=& \exp \{- {1\over \hbar} \sum '_{k,\sigma}
 \int_{-\in }^T ds \int_{-\in }^s ds'J_k^{-\sigma}(s) \nu_k (s,s')
 J_k^{-\sigma}(s')\}       \\
 &=& \int \Pi '_{k,\sigma } ({\cal D}\xi_k^{\s } P[\xi_k^{\s }])
 \exp -{i\over \hbar} \sum '_{k,\sigma} \int_{-\in }^T ds J_k^{-\s }(s)
 \xi_k^{\s }(s).
\ea
$\mid {\cal F} \mid$ is the absolute value of ${\cal F}$, containing the
kernel $\nu_k$. The phase of ${\cal F}$ contains the kernel $\mu_k$.
In the second equality, we have used a functional gaussian integral
identity, $P[\xi_k^{\s }]$ being the positive definite measure
\be
P[\xi_k^{\s }] = N\exp \{- {1\over 2\hbar} \int_{-\in }^T ds \int_{-\in }^T ds'
\xi_k^{\sigma}(s) \nu_k^{-1} (s,s') \xi_k^{\sigma}(s')\}       \\
\ee
normalized to unity. It can therefore be interpreted as a probability
distribution over the function space $\xi_k^{\s }$.

The influence functional can thus be expressed as
\ba
{\cal F}[\{Q\},\{Q'\}] &=& < \exp \{- {i\over \hbar} \sum '_{k,\s } \int_{-\in
}^T ds
J_k^{-\s }(s) [\xi_k^{\s }(s) + 2\int_{-\in }^{s} ds'\mu_k(s,s')
J_{k}^{+\s }(s')]\} >             \nn \\
&\equiv & < \exp {i\over \hbar} S_{inf} >
\ea
where $<\mbox{ }>$ denotes expectation value with respect to the joint
distribution $\Pi'_{k,\s }P[\xi_k^{\s }]$. $S_{inf}$ will be called the
stochastic influence action. We find
\ba
< \xi_k^{\s }(s) > &=& 0,     \nn \\
< \{\xi_k^{\s }(s), \xi_{k'}^{\s '}(s')\} > &=& \hbar \d _{kk'}\d _{\s \s '}
\nu_k(s,s')
\ea
where $\{\mbox{ }, \}$ denotes the anticommutator.

Substituting for $J_k^{-\s }$ and $J_k^{+\s }$ in terms of the detector
degrees of freedom $\{ Q_i \}$, the stochastic influence action $S_{inf}$ is
obtained as
\be
S_{inf} = -\sum_{i=1}^{N} \inti d\t _i \frac{dQ^- _i}{d\t _i} s_i(\t _i)
[\eta_i(\t _i) + 2\sum_{j=1}^{N} \intij d\t '_j \frac{dQ^+_j}{d\t '_j} s_j(\t
'_j)
\mt _{ij}(\t _i,\t '_j) ]
\ee
with
\be
\eta_{i}(\t _i) = e_i \sum '_{k,\s } \sqrt {2\over L} u_k^{\s }(\t _i)
\xi_k^{\s }
(t_i(\t _i)).
\ee
 From Equation ($2.29$) we see that the quantities $\mt _{ij}$, $i\neq j$
mediate
long-range interactions between the various detectors and
 the quantities $\mt _{ii}$
describe self-interaction of each detector due to its interaction
with the field. This self-interaction
typically manifests itself as a dissipative (or radiation reaction)
force in the dynamics of the detectors. We will, therefore,
refer to $\mt _{ij}$, $i\neq j$ as a ``propagation kernel'', and $\mt _{ii}$
as a ``dissipation kernel''.

We now turn to the interpretation of the quantities $\eta_i$. They appear as
source terms in the effective action of the detector system. Also, being
linear combinations of the quantities $\xi_k^{\s }$, they are stochastic
in nature. Indeed, from Equations ($2.28$) and ($2.30$) we can obtain
\ba
< \eta_i(\t _i) > &=& 0,     \nn \\
< \{\eta_i(\t _i), \eta_j(\t '_j)\}> &=& e_i e_j \sum '_{k,\s } \sum '_{k',\s
'}
u_k^{\s }(\t _i)u_{k'}^{\s '}(\t '_j) (\frac{2}{L})
< \xi_k^{\s } (t_i(\t _i)) \xi_{k'}^{\s '} (t_j(\t '_j))>   \nn \\
&=& \hbar \nt _{ij}(\t _i,\t '_j).
\ea
Thus $\nt _{ij}$ appears as a correlator of the stochastic forces $\eta_i$ and
$\eta_j$. Along a fixed trajectory, this correlation manifests as noise
in the detector dynamics. Hence we call $\nt _{ii}$ a ``noise kernel'' and
$\nt _{ij}$, $i\neq j$, a ``correlation kernel''.\footnote{The distinction
between noise and correlation is unnecessary from the point of view of the
field. `Noise', as used here, also represents free field correlations for
points on a single trajectory. However, from the point of view of each
detector, these two quantities play a different role. Hence the choice of
terminology.}

The full stochastic effective action for the $N$-detector system
is given by
\be
S_{eff} = S_{osc} + S_{inf}
\ee
We may now express this in terms of the variables $Q^+_i$ and $Q^-_i$ defined
earlier. Thus we obtain
\ba
S_{eff} &=& \sum_{i=1}^{N} \inti d\t _i [ \dot{Q^- _i} \dot{Q^+_i} -
\O _{i}^{2} Q^- _iQ^+_i
- \dot{Q^- _i}s_i(\t _i)\eta_i(\t _i) \nn \\
& &- 2 \dot{Q^- _i}
s_i(\t _i) \sum_{j=1}^{N} \intij d\t '_j \dot{Q^+_{j'}} s_j(\t '_j)
\mt _{ij}(\t _i,\t '_j) ]
\ea
where $\dot{f_i} \equiv \frac{df_i}{d\t _i}$, $\dot{f_{j'}} \equiv \frac
{df_{j}}{d\t '_j}$.

Extremizing the effective action with respect to $Q^- _i$ and setting
$Q_i = Q'_i$ at the end \cite{HPZ}, we obtain a set of coupled
equations of motion, the Langevin equations, for the system of detectors:
\be
\frac{d^2 Q_i}{d\t _i^{2}} - 2\sum_{j=1}^{N} \intij d\t '_j s_j(\t '_j)
\frac{d}{d\t _i} (s_i(\t _i)\mt _{ij}(\t _i,\t '_j) ) \frac{dQ_j}{d\t '_j}
+ \O _i^2 Q_i = \frac{d}{d\t _i}(s_i(\t _i) \eta_i(\t _i)).
\ee
Due to the back-reaction of each detector on the field, and consequently on
other detectors,
the effective dynamics of the detector system is highly non-trivial
and, as such, can be solved in closed form only for simple trajectories
or under simplifying assumptions such as ignoring the back-reaction of
certain detectors on the field. For instance, if we choose to ignore the
back-reaction of detector $i$ on the field, this can be effected by setting
$\mt _{ji}=0$, for all $j$, including $j=i$,while at the same time keeping
$\mt _{ij} \neq 0$ for $j\neq i$. The particular case $\mt _{ii}=0$ amounts to
ignoring the radiation reaction of detector $i$. This is necessary because the
radiation reaction effect arises due to a modification of the field in the
vicinity of the detector as a consequence of the back-reaction of the detector
on the field.

Of course, it is in general inconsistent to ignore the back-reaction
of a detector, as it leads to a direct violation of the symmetry ($2.17$).
As is well-known, it also leads to unphysical predictions. For example,
in the treatment of an atom on an inertial trajectory, coupled to a
quantum field, balance of vacuum fluctuations and radiation
reaction is necessary to ensure the stability of the ground state.
As explained above, ignoring back-reaction implies ignoring the
radiation reaction force. Such a treatment would render the ground
state unstable.

However, in certain cases, the quantities $\mt _{ji}$ may not contribute
to the dynamics of detector $j$, as in Section $4$ below, where the
trajectory of one detector is always outside the causal future of the
other one. Hence there is no retarded effect of one of the detectors on
the other.

Our formal treatment of the detector-field system is exact in that it
includes the full back-reaction of the detectors on the field, which
is manifested in the coupled Langevin equations of the various detectors.
The coupled equations of motion give rise to a sort of ``dynamical
correlation''
between the various detectors. Non-dynamical correlations also occur because
of the intrinsic correlations in the state of the field (Minkowski vacuum).
These
correlations are purely quantum-mechanical in origin, and they are
reflected in the correlators of the stochastic forces, $\nt _{ij}$.
Correlations between stochastic forces on different detectors
induce correlations between the coordinates $Q_{i}$ of different detectors.

As we shall show in a later section, our exact treatment makes it possible to
demonstrate the existence of generalized fluctuation-dissipation and
correlation-propagation relations
governing the
detector system.

\section{Examples}
In this and the following section, we consider some applications of the
Langevin equations
derived in the previous section to the cases of a single detector in the
Minkowski vacuum moving on an inertial trajectory, a single detector on
a uniformly accelerated trajectory, two detectors on inertial trajectories,
and the case of one detector on a uniformly accelerated trajectory
and another one on an arbitrary trajectory, functioning as a probe.

The first two examples serve to illustrate the formalism, and describe the
well-known physical effects of the dressing of a particle by the field and
the thermal Unruh noise experienced by a uniformly accelerated particle.

In the example of two inertial detectors, we introduce the notions of ``self''
and ``mutual'' impedance which govern the response of either detector. The
effect of the back-reaction of each detector on the field and consequently
on the other detector is to introduce the so-called mutual impedance in
the detector response as well as to modify the self-impedance of each detector
from its value in the absence of the other one.

In the next section we shall consider the example of one detector on a
uniformly
accelerated trajectory and
a probe, which moves along an unspecified trajectory. We switch on the
probe after it intersects the future horizon of the uniformly accelerated
detector, so that it cannot causally influence the uniformly accelerated one.
Thus the uniformly accelerated detector in this case is effectively in an
unperturbed
Unruh heat bath, and
this situation mimics most closely the RSG model. The missing terms in the RSG
analysis, which contribute to a polarization cloud
around the accelerated oscillator, but not to the energy momentum tensor,
lead to a modified noise kernel in the Langevin
equation for the probe.

In all cases, we can solve exactly for the
detector coordinates, at least in the late time limit (this limit is actually
realized at any finite time $t \gg -\infty $ when the two detectors have
been switched on forever, and corresponds to the neglect of transients in
the solutions for the detector coordinates).

\subsection{One inertial detector}

Consider the case of one detector moving on an inertial trajectory $x(\tau)=0$,
$t(\tau)=\tau$, and switched on forever ($s(\tau)=1$). The noise and
dissipation
kernels take the form
\ba
\nt (\tau,\tau') &=& \frac{e^2}{2\p }\intiz \frac{dk}{k} \cos k(\tau-\tau')  \\
\mt (\tau,\tau') &=& -\frac{e^2}{2\p }\intiz \frac{dk}{k} \sin k(\tau-\tau').
\ea
The Langevin equation becomes
\be
\frac{d^2Q}{d\tau^2} + \frac{e^2}{2} \frac{dQ}{d\tau} + \O _{0}^{2} Q =
\frac{d\eta}{d\tau}
\ee
with
\be
<\{\eta(\tau),\eta(\tau')\}> = \hbar \nt (\tau,\tau').
\ee
It will be convenient to define the dissipation constant $\ga = \frac{e^2}{4}$
. We will restrict
our attention to the underdamped case ($\ga \leq \O_0$ ).

Introducing the Fourier transform
\ba
Q(\tau) &=& \intin d\o e^{i\o \t } \Qt (\o )        \nn  \\
\Qt(\o ) &=& \frac{1}{2\p }\intin d\o e^{-i\o \t } Q(\tau )
\ea
and similarly for $\eta(\t )$, we obtain
\be
\Qt (\o ) = \cw \etat (\o )
\ee
with the impedance function $\cw $ defined as
\be
\cw  = i\o (-\o ^2 + \O _0^2 + 2i\o \ga )^{-1}.
\ee
In the above solution for the detector coordinate in frequency space, it should
be noted that transients have already been neglected. Transient terms
correspond
to delta functions in frequency space, the coefficients of these delta
functions
being determined by the initial conditions. For the complete solution these
terms should be added to the right hand side of Equation ($3.6$).
We may thus obtain
\be
<\{ \Qt (\o ), \Qt (\o ')\} > = \cw \cwp <\{ \etat (\o ), \etat (\o ')\} >
\ee
where
\ba
<\{\etat (\o ), \etat (\o ')\}> &=& \frac{1}{4\p ^2} \intin d\t \intin d\t '
e^{-i\o \t } e^{-i\o '\t '} <\{\eta(\t ), \eta(\t ')\}>      \nn \\
&=& \hbar \frac{\ga }{\p \o }\d (\o +\o ') [\q (\o ) - \q (-\o )].
\ea
We can therefore obtain the correlator of $Q(\t )$ and $Q(\t ')$, as
\ba
<\{ Q(\t ), Q(\t ')\}> &=& \intin d\o \intin d\o ' e^{i\o \t }e^{i\o '\t '}
\cw \cwp <\{\etat (\o ), \etat (\o ')\}>    \nn \\
&=& \frac{2\hbar \ga}{\p } \intiz \frac{dk}{k} \mid \ck \mid^2 \cos k(\t -\t
').
\ea

\subsection{One accelerated detector: Unruh effect}

In the case of an accelerated detector moving on the trajectory
$x(\t ) = a^{-1}\cosh a\t $, $t(\t ) = a^{-1}\sinh a\t $, and $s(\t ) = 1$
($\t $ being the proper time along the accelerated trajectory),
the noise and dissipation kernels take the form:
\ba
\nt (\tau,\tau') &=& \frac{e^2}{2\p }\intiz \frac{dk}{k} \cos \frac{k}{a}
(\sinh a\t -\sinh a\t ') \cos \frac{k}{a} (\cosh a\t - \cosh a\t ')\\
\mt (\tau,\tau') &=& -\frac{e^2}{2\p }\intiz \frac{dk}{k} \sin \frac{k}{a}
(\sinh a\t -\sinh a\t ') \cos \frac{k}{a} (\cosh a\t - \cosh a\t ').
\ea
These kernels can be decomposed into advanced and retarded parts, by
writing, for example
\be
\nt ^a + \nt ^r = \frac{e^2}{4\p }\intiz \frac{dk}{k} \Bigl[\cos \frac{k}{a}
(e^{a\t} - e^{a\t '}) + \cos \frac{k}{a}
(e^{-a\t} - e^{-a\t '})\Bigr]\;.
\ee
We can then use the changes of variables $k \to \frac{k}{2}
e^{\pm\frac{a}{2}(\t -\t ')}$ to obtain
\be
\nt ^a = \nt^r =  \frac{e^2}{4\p }\intiz \frac{dk}{k}\cos \Bigl(\frac{k}{a}
\sinh\frac{a}{2}(\t -\t ')\Bigr)\;,
\ee
showing that the noise felt by the accelerating detector is isotropic.

One can also make a similar simplification for the kernel $\mt$.
These expressions can then be further simplified \cite{Anglin,HM2}
by means of the integral transform \cite{GR1}
\be
e^{ik\sinh b} = \frac{2}{\p } \intiz d\a \cos (b\a +i\p \a /2) K_{i\a }(k)
\ee
and the formula \cite{GR2}
\be
\intiz \frac{dk}{k} K_{i\a}(k) = \frac{\p }{2\a \sinh{\p \a \over 2}}\;,
\ee
where $K_{i\a }(a)$ is a Bessel function of imaginary argument,
to yield
\ba
\nt (\tau,\tau') &=& \frac{e^2}{2\p }\intiz \frac{dk}{k} \coth (\frac{\p k}{a})
\cos k(\tau-\tau')  \\
\mt (\tau,\tau') &=& -\frac{e^2}{2\p }\intiz \frac{dk}{k} \sin k(\tau-\tau').
\ea
The noise experienced by the detector is thus stationary and the factor
$\coth (\frac{\p k}{a})$ in the noise kernel shows that it
 is also thermal, at the Unruh temperature
$k_{B}T = \frac{\hbar a}{2\p }$ (we have chosen units such that $c=1$).

The dissipation kernel
 remains identical to that of the inertial detector. Based on the property that
 the two-point function of a free field on an accelerated trajectory evaluated
 in the Minkowski vacuum state is
identical to the two-point function on an inertial trajectory evaluated in a
thermal state at the Unruh temperature, this fact can be explained as follows:
The dissipation kernel is proportional to the commutator of the free quantum
field
evaluated in whatever state the field is in. However, the commutator of a free
field for any two points is just a $c$-number, hence its expectation value
is independent of the state of the field. In particular, it does not
distinguish between a zero temperature and a thermal state. So the dissipation
kernel is identical to that in the inertial case. The anticommutator is,
however,
an operator whose expectation value depends on the state of the field, and
therefore shows the familiar departure from the inertial case.

The Langevin equation for the detector coordinate is
\be
\frac{d^2Q}{d\tau^2} + \frac{e^2}{2} \frac{dQ}{d\tau} + \O _{0}^{2} Q =
\frac{d\eta}{d\tau}
\ee
with
\be
<\{\eta(\tau),\eta(\tau')\}> = \hbar \nt (\tau,\tau').
\ee
Similar to the inertial detector case, we find
\be
\Qt (\o ) = \cw \etat (\o ).
\ee
Also
\ba
<\{\etat (\o ), \etat (\o ')\}> &=& \frac{1}{4\p ^2} \intin d\t \intin d\t '
e^{-i\o \t } e^{-i\o '\t '} <\{\eta(\t ), \eta(\t ')\}>      \nn \\
&=& \hbar \frac{\ga }{\p \o } \coth (\frac{\p \o }{a}) \d (\o +\o ').
\ea
Combining the two equations,
\ba
<\{ Q(\t ), Q(\t ')\}> &=& \intin d\o \intin d\o ' e^{i\o \t }e^{i\o '\t '}
\cw \cwp <\{\etat (\o ), \etat (\o ')\}>    \nn \\
&=& \frac{2\hbar \ga}{\p } \intiz \frac{dk}{k} \mid \ck \mid^2 \coth (\frac
{\p k}{a}) \cos k(\t -\t ')
\ea
with the impedance function $\ck $ as defined in the inertial case.

\subsection{Two inertial detectors: self and mutual impedance}

We now consider the case of two detectors moving on the inertial
trajectories $x_1(\t _1)=-x_0/2$, $x_2(\t _2)=x_0/2$ and $t_1(\t _1)=t_2(\t _2)
=\t $, coupled to a scalar field initially in the Minkowski
 vacuum state, with coupling constants $e_{1,2}$. They are separated by a
fixed coordinate distance $x_0$. As before, we will assume that both detectors
have been forever switched on, i.e. $s_i(\t )=1$, $i=1,2$.

It will be convenient to express the noise, dissipation, correlation
and propagation
kernels as the real and imaginary parts of the functions $Z_{ij}$ defined
earlier. Then, for the two-detector system, we obtain
\ba
Z_{11}(\t ,\t ') &=& \frac{e_1^2}{2\p }\intiz \frac{dk}{k} e^{-ik(\t -\t ')} \\
Z_{22}(\t ,\t ') &=& \frac{e_2^2}{2\p }\intiz \frac{dk}{k} e^{-ik(\t -\t ')} \\
Z_{12}(\t ,\t ') = Z_{21}(\t ,\t ') &=& \frac{e_1 e_2}{2\p }\intiz \frac{dk}{k}
e^{-ik(\t -\t ')} \cos kx_0 .
\ea
The coupled Langevin equations for the system are
\ba
\frac{d^2Q_1}{d\tau^2} + \frac{e_1^2}{2} \frac{dQ_1}{d\tau} + \frac{e_1e_2}{2}
\frac{dQ_2}{d\t }\mid_{\t -x_0} + \O _{1}^{2} Q_1 &=& \frac{d\eta_1}{d\tau}  \\
\frac{d^2Q_2}{d\tau^2} + \frac{e_2^2}{2} \frac{dQ_2}{d\tau} + \frac{e_1e_2}{2}
\frac{dQ_1}{d\t }\mid_{\t -x_0} + \O _{2}^{2} Q_2 &=& \frac{d\eta_2}{d\tau}
\ea
where $\t -x_0$ is the retarded time between the two trajectories, and
\be
<\{\eta_{i}(\t ),\eta_{j}(\t ')\}> = \hbar \nt _{ij}(\t -\t ').
\ee

As before, we define $\ga _{1,2}=\frac{e_{1,2}^2}{4}$, and introduce Fourier
transforms to obtain the corresponding equations in frequency space. Then we
obtain
\ba
\Qot (\o ) &=& \cwo \etaot (\o ) - \frac{e_1e_2}{2}e^{-i\o x_0} \cwo \Qtt (\o )
\\
\Qtt (\o ) &=& \cwt \etatt (\o ) - \frac{e_1e_2}{2}e^{-i\o x_0} \cwt \Qot (\o )
\ea
where
\be
\cw ^{(1),(2)} = i\o (-\o ^2 + \O _{1,2}^2 + 2i\o \ga _{1,2})^{-1}.
\ee
The functions $\cw ^{(1),(2)}$ are, of course, what the impedance of
each detector would be in the absence of the other one. However, the
effect of introducing a second detector is, as we shall see, to modify
the ``self - impedance'' of each detector as well as introduce a ``mutual
impedance'' which describes, for instance, the response of detector $1$ to
the force $\etatt $.

Indeed, plugging the equation for $\Qot $ in the equation for $\Qtt $, we have
\be
\Qtt (\o ) = L_{22}(\o ) \etatt (\o ) + L_{21}(\o ) \etaot (\o )
\ee
where $L_{22}$ is the modified self - impedance of detector $2$ due to the
presence of detector $1$, and $L_{21}$
is the mutual impedance:
\ba
L_{22}(\o ) &=& \cwt (1 - 4\ga _1\ga _2 e^{-2i\o x_0}\cwo \cwt )^{-1} \nn \\
L_{21}(\o ) &=& -2\sqrt{\ga _1\ga _2} e^{-i\o x_0}\cwt \cwo
(1 - 4\ga _1\ga _2 e^{-2i\o x_0}\cwo \cwt )^{-1}.
\ea
The impedances $L_{11}$ and $L_{12}$ and the corresponding equation for $\Qot $
are obtained by an interchange of indices $1$ and $2$ in the above
equations.

We note the symmetry
\be
L_{21} = L_{12}.
\ee
The correlator $<\{Q_i(\o ), Q_j(\o ')\}>$, $i,j=1,2$ is therefore obtained
from
equation ($3.32$) and its counterpart, as
\be
<\{Q_i(\o ), Q_j(\o ')\}> = \sum_{\a =1}^{2} \sum_{\b =1}^{2} L_{i\a }(\o )
L_{j\b }(\o ') <\{ \tilde{\eta}_{\a }(\o ), \tilde{\eta}_{\b }(\o ')\}>.
\ee
The above equation is to be viewed as a generalization of ($3.8$) to the
two-detector case.

Suppose we now wish to solve for the correlator of $Q_2$. Then, taking Fourier
 transforms as before and simplifying,
\ba
& &<\{ Q_2(\t ), Q_2(\t ')\}> = \intin d\o \intin d\o ' e^{i\o \t }e^{i\o '\t
'}
<\{ \Qtt (\o ), \Qtt (\o ')\}>  \\ \nn
&=& \frac{2\hbar \ga _2}{\p } \intiz \frac{dk}{k} \mid L_{22}(k)\mid^2 [1 + \ga
_1(4 \ga _1 \mid \ck ^{(1)} \mid^2 - \ck ^{(1)} - \ck ^{(1)\ast}) \\ \nn
& &- \ga _1 (\ck
^{(1)} e^{-2ikx_0} + \ck ^{(1)\ast} e^{2ikx_0}) ] .
\ea
The second term in the square brackets vanishes as a consequence of the
identity $\ck ^{(1)} + \ck ^{(1)\ast} = 4\ga _1\mid \ck ^{(1)} \mid^2$, which
is
a form of the fluctuation dissipation relation for detector $1$. The
remaining terms simplify to yield
\ba
& &<\{ Q_2(\t ), Q_2(\t ')\}> =  \\ \nn
& & \frac{2\hbar \ga _2}{\p } \intiz \frac{dk}{k} \mid L_{22}(k)\mid^2 [1 -
2\ga
 _1 \mid \cwo \mid^2 (2\ga _1\cos 2kx_0 + \frac{\O _1^2 - k^2}{k} \sin 2kx_0)]
 \times \\ \nn
& & \cos k(\t -\t ').
\ea

As before, the correlator of $Q_1$ is obtained by interchanging the
indices $1$ and $2$ in the above equation.
\section{Uniformly accelerated detector and probe: advanced and retarded
noises}

In this section, we first consider two detectors on arbitrary trajectories,
with the following constraints: $a)$ both trajectories are everywhere timelike,
$b)$ one of the trajectories possesses past and future event horizons, which
are chosen to be the null lines $v=0$ and $u=0$ respectively, $c)$ the detector
on the other trajectory is switched on at $u=0$ and this trajectory does not
possess a future horizon.

Because of constraint $c)$, the second detector cannot causally influence
the first one, and thus it functions as a probe in the field modified by
the first detector.

Later in the analysis, we shall specify the trajectory of the detector
with horizons as being a uniformly accelerated one. We shall continue to
assume that the probe cannot causally influence the uniformly accelerated
detector by means of the switching condition. If it were allowed to do so,
this would lead to a deviation of the noise experienced by the uniformly
accelerated detector from the precise thermal form.

We will label the detector with horizons as detector $1$ and the probe
 as detector $2$. The switching condition $s_2(\t _2)=\q (u_2(\t _2))$ for
 the probe leads to a closed Langevin equation for detector $1$:
\be
\frac{d^2Q_1}{d\tau_1^2} + \frac{e_1^2}{2} \frac{dQ_1}{d\tau_1} + \O _{1}^{2}
Q_1 = \frac{d\eta_1}{d\tau_1}.
\ee
This is just a consequence of the fact that the trajectory of detector $1$
lies outside the causal future of the probe. The arguments which lead to
the above local form of dissipation or radiation reaction for a general
timelike
trajectory are outlined in the next section.

Introducing Fourier transforms and the impedance functions $\cw ^{(1),(2)}$
as defined earlier, we have,
\be
\Qot (\o ) = \cwo \etaot (\o ).
\ee

Consider now the Langevin equation for detector $2$:
\ba
& &\frac{d^2Q_2}{d\t _2^2} - 2\int_{-\infty}^{\t _2} d\t '_2 \frac{d}{d\t _2}
(\q (u_2(\t _2))\mt _{22}(\t _2,\t '_2)) \frac{dQ_2}{d\t '_2}  \nn \\
& & - 2\int_{-\infty}^{t_1^{-1}(t_2(\t _2))}
d\t '_1 \frac{d}{d\t _2}(\q (u_2(\t _2))\mt _{21}(\t _2,\t '_1))\frac{dQ_1}
{d\t '_1} + \O _2^2Q_2 = \frac{d}{d\t _2}(\q (v_2(\t _2))\eta_2(\t _2)).
\ea
We find
\ba
\frac{d}{d\t _2}(\mt _{22}(\t _2,\t '_2)) &=& -2\ga _2 \d (\t _2-\t '_2) \\
\frac{d}{d\t _2}(\mt _{21}(\t _2,\t '_1)) &=& -\sqrt{\ga _1\ga _2}
[\frac{dv_2}{d\t _2}
\d (v_2(\t _2)-v_1(\t '_1)) + \frac{du_2}{d\t _2}\d (u_2(\t _2)-u_1(\t '_1))]
\ea
where $\ga _{1,2}$ are defined as in the two inertial detector case.
The second term in ($4.5$) vanishes identically because $u_2(\t _2)>0$ and
$u_1(\t _1)<0$ ($u=0$ is a future horizon for detector $1$). Since $v=0$ is
a past horizon for detector $1$, we have $v_1(\t '_1)>0$ and the first
term simplifies to yield
\be
\frac{d}{d\t _2}(\mt _{21}(\t _1,\t '_2)) = -\sqrt{\ga _1\ga _2} \frac{d\t _R}{
d\t _2}\d (\t _R-\t '_1) \q (v_2(\t _2))
\ee
where we have defined the retarded time $\t _R = v_1^{-1}(v_2(\t _2))$. This
is well-defined since it occurs only in expressions in which $v_2(\t _2)>0$.

Thus we obtain the dynamical equation for the probe, which depends, as
expected,
on $Q_1$:
\be
\frac{d^2Q_2}{d\tau_2^2} + \frac{e_2^2}{2} \frac{dQ_2}{d\tau_2} + \O _{2}^{2}
Q_2 = \frac{d\eta_2}{d\tau_2} - \frac{e_1e_2}
{2}\frac{d\t _R}{d\t _2}\frac{dQ_1}{d\t }\mid_{\t _R}, ~~~ v_2(\t _2)>0.
\ee
Consider the quantity
\be
F(\t _2) = \frac{d\eta_2}{d\tau_2} - \frac{e_1e_2}
{2}\frac{d\t _R}{d\t _2}\frac{dQ_1}{d\t }\mid_{\t _R}
\ee
which is a source term in the equation of motion for $Q_2$. The first part
of $F$ is the usual stochastic force arising out of the fluctuations of the
field in the vicinity of detector $2$, while the second part is the
retarded force due to detector $1$. RSG correctly point out that these two
forces are correlated. In the context of our formalism, these correlations
are embodied in the correlation kernels $\nt _{21}$ and $\nt _{12}$.

Using the relation ($4.2$), we obtain
\be
F(\t _2) = \frac{d\eta_2}{d\tau_2} - i\frac{e_1e_2}{4\p }\frac{d\t _R}{d\t _2}
\intin d\o \o e^{i\o \t _R}
\cwo \intin ds \eta_1(s) e^{-i\o s}.
\ee
Consider the correlator of $F$ with the correlator of $\eta_2$ subtracted
out. We have:
\ba
& &<\{F(\t _2), F(\t '_2)\}> - \frac{d}{d\t _2} \frac{d}{d\t '_2}<\{\eta_2(\t
_2), \eta_2(\t '_2)\}> =     \\ \nn
& & -\frac{i\hbar \sqrt{\ga _1\ga _2}}{\p } \intin d\o \o \cwo \intin ds
e^{-i\o
 s}[\frac{d\t '_R}{d\t '_2}e^{i\o \t '_R} \frac{d}{d\t _2} \nt _{21}(\t _2,s) +
\frac{d\t _R}{d\t _2}e^{i\o \t _R} \frac{d}{d\t '_2} \nt _{12}(s,\t '_2)] \\
\nn
& & -\frac{\hbar \ga _1\ga _2}{\p ^2} \frac{d\t _R}{d\t _2}\frac{d\t '_R}{d\t '
_2}\intin d\o \intin d\o ' \o \o 'e^{i\o \t _R}e^{i\o '\t '_R} \cwo \cwop
\intin ds \intin ds' e^{-i\o s}
e^{-i\o 's'} \nt _{11}(s,s') .
\ea
The kernels $\nt _{21}$ and $\nt _{12}$ separate into advanced and retarded
parts. For the advanced parts,
\ba
\nt _{21}^{a}(\t _2,s) &=& \frac{e_1e_2}{4\p }\intiz \frac{dk}{k} \cos k(v_2(\t
 _2) - v_1(s))             \nn \\
&=& \frac{e_1e_2}{4\p }\intiz \frac{dk}{k} \cos k(v_1(\t _R) - v_1(s))
= \frac{e_2}{e_1} \nt _{11}^{a}(\t _R,s)
\ea
and similarly
\be
\nt _{12}^{a}(s,\t '_2) = \frac{e_2}{e_1} \nt _{11}^{a}(s,\t '_R).
\ee
The advanced parts of the correlations can therefore be constructed from
the advanced part of the noise along the trajectory of detector $1$.
With this simplification, we obtain
\ba
& &<\{F(\t _2), F(\t '_2)\}> - \frac{d}{d\t _2} \frac{d}{d\t '_2}<\{\eta_2(\t
_2), \eta_2(\t '_2)\}> =     \nn \\
& & -\frac{i\hbar \sqrt{\ga _1\ga _2}}{\p } \frac{d\t _R}{d\t _2}\frac{d\t '_R}
{d\t '_2}[\intin d\o \o \cwo \intin ds e^{-i\o s}
\{ e^{i\o \t '_R} \frac{d}{d\t _R} \nt _{11}^{a}(\t _R,s) +
e^{i\o \t _R} \frac{d}{d\t '_R} \nt _{11}^{a}(s,\t '_R)\}  \nn \\
& & -i\frac{\ga _1}{\p } \intin d\o \intin d\o ' \o \o 'e^{i\o \t _R}
e^{i\o '\t '_R} \cwo \cwop
\intin ds \intin ds' e^{-i\o s}
e^{-i\o 's'} \nt _{11}(s,s')]  \nn \\
& & + (r.p.)
\ea
where $(r.p.)$ denotes the retarded part:
\ba
& & (r.p.) =    \\ \nn
& & -\frac{i\hbar \sqrt{\ga _1\ga _2}}{\p } \intin d\o \o \cwo \intin ds
e^{-i\o
 s}[\frac{d\t '_R}{d\t '_2} e^{i\o \t '_R} \frac{d}{d\t _2} \nt _{21}^{r}(\t
_2,
 s) +
\frac{d\t _R}{d\t _2}e^{i\o \t _R} \frac{d}{d\t '_2} \nt _{12}^{r}(s,\t '_2)] .
\ea

At this point, we specialize to the case when detector $1$ is uniformly
accelerated. Then we have
\be
v_1(\t _1) = a^{-1} e^{a\t _1} ; ~~~ u_1(\t _1) = -a^{-1} e^{-a\t _1}.
\ee
As shown in subsection 3.2, the noise $\nt _{11}$ is thermal and isotropic.
The retarded time $\t _R = a^{-1}\ln (av_2(\t _2))$. We may substitute for
$\nt _{11}$ in equation ($4.13$) and carry out the integrations over $s$ and
$s'$
to obtain
\ba
& &<\{F(\t _2), F(\t '_2)\}> - \frac{d}{d\t _2} \frac{d}{d\t '_2}<\{\eta_2(\t
_2), \eta_2(\t '_2)\}> =     \\ \nn
& & -\frac{2\hbar \ga _1\ga _2}{\p } \frac{d\t _R}{d\t _2}\frac{d\t '_R}
{d\t '_2}\intiz dk k \coth\frac{\p k}{a} \cos k(\t _R-\t '_R) (\ck ^{(1)} +
\ck ^{(1)\ast} - 4\ga _1 \mid \ck ^{(1)} \mid^2) + (r.p.).
\ea
The first term in the above expression vanishes as a consequence of the
identity $\ck ^{(1)} + \ck ^{(1)\ast} = 4\ga _1\mid \ck ^{(1)} \mid^2$,
mentioned earlier. The only contribution to the excitation of the probe
is therefore from the retarded parts of the correlations $\nt _{12}$ and
$\nt _{21}$. This asymmetry between retarded and advanced parts is really a
consequence of the choice of retarded boundary conditions in the formulation
of the problem (the states of detector and field are assumed to be uncorrelated
at past infinity) and the switching process at $u_2=0$. The vanishing of the
first term in the above expression is a generalization of the cancellation
obtained by RSG for a probe moving along an inertial trajectory.
In order to study the retarded contribution in greater detail, it is
desirable to simplify the correlations $\nt _{12}^{r}$ and $\nt _{21}^{r}$.
The functions $Z_{12}^{r}$ and $Z_{21}^{r}$ take the form
\ba
Z_{12}^{r}(\t _1, \t '_2) &=& \frac{e_1e_2}{4\p } \intiz \frac{dk}{k} e^{
\frac{ik}{a}(e^{-a\t _1} + au_2(\t '_2))} \\
Z_{21}^{r}(\t _2, \t '_1) &=& \frac{e_1e_2}{4\p } \intiz \frac{dk}{k} e^{
-\frac{ik}{a}(e^{-a\t '_1} + au_2(\t _2))}.
\ea
Introducing the Fourier transforms
\ba
e^{\frac{ik}{a}e^{-a\t _1}} &=& \frac{1}{2\p a} \intin d\o e^{-i\o \t _1} \G
(-\frac{i\o }{a}) (\frac{k}{a})^{\frac{i\o }{a}} e^{\frac{\p \o }{2a}} ,~~~ k>0
\nn \\
e^{-\frac{ik}{a}e^{-a\t _1}} &=& \frac{1}{2\p a} \intin d\o e^{i\o \t _1} \G
(\frac{i\o }{a}) (\frac{k}{a})^{-\frac{i\o }{a}} e^{\frac{\p \o }{2a}} ,~~~ k>0
\ea
and substituting in ($4.19$), we obtain
\be
Z_{12}^{r} (\t _1,\t '_2) = \frac{e_1e_2}{8\p ^2 a} \intin d\o e^{-i\o \t _1}
e^{\frac{\p \o }{2a}} \G (-\frac{i\o }{a}) \intiz \frac{dk}{k}(\frac{k}{a})^
{\frac{i\o }{a}} e^{iku_2(\t '_2)}.
\ee
The integral over $k$ can be evaluated in terms of gamma functions. Explicitly
making use of the fact that $u_2(\t '_2)>0$, and
simplifying,
\be
Z_{12}^{r} (\tau_1,\tau'_2) = \frac{e_1e_2}{4\p }\intiz \frac{d\o }{\o }
 \frac{\cos \frac{\o }{a}(a\t _1+\ln au_2(\t '_2))}{\sinh \frac{\p \o }{a}}.
\ee
Similarly,
\be
Z_{21}^{r} (\tau_2,\tau'_1) = \frac{e_1e_2}{4\p }\intiz \frac{d\o }{\o }
 \frac{\cos \frac{\o }{a}(a\t '_1+\ln au_2(\t _2))}{\sinh \frac{\p \o }{a}}.
\ee
Therefore there is no contribution from the imaginary parts of $Z_{12}^{r}$ and
$Z_{21}^{r}$ for $u_2(\t '_2)>0$, as obtained earlier (equation ($4.5$) leading
to ($4.6$)).

Differentiating the above expressions with respect to $\t '_2$ and $\t _2$,
and substituting in the expression for $(r.p.)$, one obtains, after
carrying out the integration over $s$,
\ba
& &<\{F(\t _2), F(\t '_2)\}> - \frac{d}{d\t _2} \frac{d}{d\t '_2}<\{\eta_2(\t
_2), \eta_2(\t '_2)\}> \equiv (r.p.) \\ \nn
& &= \frac{2\hbar \ga _1\ga _2}{\p } \intin \frac{dk k}{\sinh \pka}
 \ck ^{(1)} [(a^2 u_2(\t '_2)v_2(\t _2))^{\ika -1}\frac{du_2}{d\t '_2}
 \frac{dv_2}{d\t _2} + (a^2 u_2(\t _2)v_2(\t '_2))^{\ika -1}\frac{du_2}{d\t _2}
\frac{dv_2}{d\t '_2}].
\ea
The coincidence limit of the above expression yields the fluctuations of the
random force acting on the probe: defining $\d F(\t )=F(\t )-\frac{d}{d\t }\eta
_2(\t )$, we obtain
\be
<\d F^2(\t _2)> = \frac{2\hbar \ga _1\ga _2}{\p } \frac{du_2}{d\t _2}
 \frac{dv_2}{d\t _2}\intin \frac{dk k}{\sinh \pka}
 \ck ^{(1)} (a^2 u_2(\t _2)v_2(\t _2))^{\ika -1}.
\ee
The fluctuations are thus suppressed in the limit of large $u_2v_2 = t_2^2
-x_2^2$. For a probe trajectory without horizons, this is the limit in which
the probe trajectory approaches future timelike infinity, which verifies
that the effect of the accelerated oscillator on the field is ascribed to
polarization rather than radiation (see also \cite{MPB}). A radiation field
is expected to persist at future infinity.

Let us now turn to the question of the response of the probe. To obtain this,
we will need to specify a particular form of trajectory for the probe as well.
We will consider the simple inertial trajectory $x_2(\t _2)=0$, $t_2(\t _2)
=\t _2$, switched on at $\t _2=0$.
Then equation ($4.23$) gives
\ba
& &<\{F(\t _2), F(\t '_2)\}> - \frac{d}{d\t _2} \frac{d}{d\t '_2}<\{\eta_2(\t
_2), \eta_2(\t '_2)\}> = \\ \nn
& &\frac{2\hbar \ga _1\ga _2}{\p } \intin \frac{dk k}{\sinh \pka}
 \ck ^{(1)} (a\t _2)^{\ika -1}(a\t '_2)^{\ika -1}.
\ea

Owing to the switching process at $\t _2=0$, the relation between $\Qtt $ and
$\Ft $
(the Fourier transforms of $Q_2$ and $F$) is a non-local one in frequency
space, because of transient effects. However, if we restrict our attention to
the late time behavior of detector $2$, we obtain from equation ($4.7$) a local
relation of the form
\be
\Qtt (\o ) = \frac{\cwt }{i\o }\Ft (\o )
\ee
where
\be
\Ft (\o ) = \frac{1}{2\p }\intiz d\t e^{-i\o \t }F(\t ).
\ee
In the above expression, the lower limit of the $\t $ integration is zero,
corresponding to the step function $\q (\t )$ multiplying $F$ which
enforces the switching condition.

The correlator of $\Qtt $ is therefore given by
\be
<\{ \Qtt (\o ), \Qtt (\o ')\}> = -\frac{\cwt \cwtp}{\o \o '}  <\{ \Ft (\o ),
\Ft (\o ')\}>.
\ee
We have already obtained the difference of the correlator of $F$ from its
value in the absence of the accelerating detector, $1$. Thus we have
\ba
& &<\{ \Qtt (\o ), \Qtt (\o ')\}> - <\{ \Qtt ^{(0)}(\o ), \Qtt ^{(0)}(\o ')\}>
=  \\
& &-\frac{\cwt \cwtp }{4\p ^2\o \o '} \intiz d\t \intiz d\t ' e^{-i\o \t
}e^{-i\o '\t '
}[<\{F(\t ), F(\t ')\}> - \frac{d}{d\t } \frac{d}{d\t '}<\{\eta_2(\t ), \eta_2
(\t ')\}>] \nn
\ea
where the superscript $(0)$ on $Q_2$ refers to its value in the absence
of the accelerating detector. Performing the integrations over $\t $ and
$\t '$, we obtain
\ba
& &<\{ \Qtt (\o ), \Qtt (\o ')\}> - <\{ \Qtt ^{(0)}(\o ), \Qtt ^{(0)}(\o ')\}>
= \nn  \\
& & - \frac{\hbar \ga _1\ga _2}{2\p ^3a^4} \frac{\cwt \cwtp }{\o \o '}
\intin \frac{dk k}{\sinh \pka} \ck ^{(1)} \G ^2(\ika ) \mid \frac{\o \o '}{a^2}
\mid^{-\ika -1} (e^{\pkta }\q (\o ) + e^{-\pkta }\q (-\o )) \times \nn \\
& &(e^{\pkta }\q (\o ') + e^{-\pkta }\q (-\o ')).
\ea
The step functions which distinguish positive and negative frequencies
in the above expression are an artefact of the switching process.

\section{Fluctuation-dissipation and correlation-propagation relations}

In this section we construct the fluctuation-dissipation relations for the
detector system and extend this construction to obtain a new set of relations,
which we call the correlation-propagation
relations for trajectories without event horizons. These relations are
a simple consequence of the analytic properties
of the massless free field two-point function. We also discuss these relations
in the context of the model of a uniformly accelerated detector and probe.

Consider first the fluctuation dissipation relation for a quantum
Brownian particle in a heat bath \cite{HPZ}. This can be expressed
as a linear, non-local relation between the noise $\nt (s)$ and dissipation
$\mt (s)$
kernels. Defining $\gt $ by
\be
\mt (s) = \frac{d\gt }{ds}(s),
\ee
the finite temperature fluctuation dissipation relation is
\be
\nt (s) = \intin ds' K(s-s') \gt (s')
\ee
where
\be
K(s-s') = \intiz \frac{kdk}{\p } \coth (\frac{\b k\hbar}{2})\cos k(s-s')
\ee
is a universal kernel, independent of the spectral density of the bath.
In particular, the kernel $K$ is independent of the coupling constant $e$. Such
 a fluctuation-dissipation relation holds for the
uniformly accelerated detector (with temperature given by the Unruh
temperature) and the inertial detector (with zero temperature).
It was derived in \cite{HPZ} in the context of a quantum Brownian
model
with bilinear coupling between bath (field) and particle (detector). In such
a model $\gt $ is indeed the quantity which characterizes dissipation in the
effective Langevin equation for the particle.
\\

In the context of the minimally coupled model, however, we find it suitable to
define $\gt $ as
\be
\gt (s) = -\frac{d}{ds} \mt (s)
\ee
as this is the quantity which directly appears in the dissipative term of the
Langevin equation derived above. The fluctuation-dissipation relation then
takes
the form ($5.3$) with $K$ defined as
\be
K(s-s') = \intiz \frac{dk}{\p k} \coth (\frac{\b k\hbar}{2})\cos k(s-s').
\ee
An important
aspect of either form of the fluctuation-dissipation relation is that the
noise and dissipation kernels, and consequently $K$, are stationary, i.e. they
are functions of $s-s'$ alone.

We wish to investigate whether a suitable generalization of the above
relation holds for the full $N$-detector system. To this end, we
 assume that the detector trajectories are everywhere timelike and consider
 first only the kernels $Z_{ii}$ as they characterize noise
and dissipation in the dynamics of the detectors $i$. We also assume that the
detectors are switched on forever,
thus excluding transient effects due to the switching process. Using advanced
and retarded null coordinates introduced earlier, we define
\ba
\gt _{ii} (\t _i, \t '_i) &=& -\frac{d}{d\t _i} \mt _{ii}(\t _i, \t '_i) \nn \\
&=&\frac{e_i^2}{4}[\d (v_i(\t _i) - v_i(\t '_i))\frac{dv_i}{d\t _i} + \d
(u_i(\t _i) - u_i(\t '_i))\frac{du_i}{d\t _i}]    \nn \\
&=& \gt _{ii}^{a}(\t _i,\t '_i) + \gt _{ii}^{r}(\t _i,\t '_i),
\ea
denoting the advanced and retarded parts of the kernel $\gt $.

The timelike property of the trajectories implies that $\mid\frac{dx_i}
{dt _i}\mid<1$. Together with the fact that $t_i(\t _i)$ are increasing
functions
of $\t _i$, this implies that $\frac{du_i}{d\t _i}$ and $\frac{dv_i}{d\t _i}$
are necessarily positive. It also implies that the functions $u_i(\t _i)$ and
$v_i(\t _i)$ have unique inverses, if they exist. This can be proved by way of
contradiction: assume that $u_i(\t _i)=u_i(\t '_i)$ for some $\t _i \neq
\t '_i$. Then we have $x_i(\t _i)-x_i(\t '_i)=t_i(\t _i)-t_i(\t '_i)$, which
means that the points $\t _i$ and $\t '_i$ have light-like separation. This
contradicts the fact that the trajectory is everywhere timelike. The uniqueness
of $v_i^{-1}$ is shown in the same way.

These two properties lead to the following simplification in the expression
for $\gt _{ii}$:
\be
\gt _{ii}(\t _i,\t '_i) = \frac{e_i^2}{2} \d (\t _i-\t '_i).
\ee
Thus we see that, for an arbitrary trajectory, the dissipation or
radiation reaction kernel has the same form and is always local. This fact has
been used in obtaining the dissipative term in the equations of motion for
the accelerated detector and probe ($4.1$ and $4.7$).

The fluctuation-dissipation relation now follows in a straightforward
manner:
\be
\nt _{ii}(\t _i,\t '_i) = \intin ds K_i(\t _i,s) \gt _{ii}(s, \t '_i)
\ee
where
\ba
K_i(\t _i,s) &=& K_i^{a}(\t _i,s) + K_i^{r}(\t _i,s)  \nn \\
&=&\intiz \frac{dk}{2\p k} [\cos k(v_i(\t _i) - v_i(s)) + \cos k(u_i(\t _i)
- u_i(s))].
\ea

We now ask whether a similar relation holds between the real and imaginary
parts of $Z_{ij}$, $i\neq j$. This would not be a fluctuation dissipation
relation in the usual sense, as the real part of $Z_{ij}$ describes
correlations
of the field between points on different trajectories rather than fluctuations,
and its imaginary part describes the propagation of radiation between one
detector and the
other, rather than dissipation. We will call such relations ``correlation-
propagation'' relations.

If points on different trajectories have
space like separations, the relevant $\gt _{ij}$ (defined as $-\frac{d\mt
_{ij}}
{d\t _i}$) will vanish as a consequence
of the vanishing of the commutator of a free field for points at spacelike
separations. This is simply an expression of causality in the detector
dynamics.
However, the corresponding correlation $\nt _{ij}$ need not vanish, and hence
there cannot be a general relation between these two kernels. Such a situation
is realized most clearly, for example, in the case of two uniformly
accelerating
detectors, one in the right and the other in the left Rindler wedge. The
trajectories, although individually timelike, are spacelike separated
everywhere. The
corresponding $\gt _{12}$ and $\gt _{21}$ will therefore vanish identically.
However, $\nt _{12}$ and $\nt _{21}$ will remain non-zero, reflecting the
highly correlated nature of the Minkowski vacuum state.

If, however, none of the detector trajectories possess past or future horizons
(in Minkowski space this is true in particular for geodesic trajectories, but
not $only$ for geodesic trajectories), then
 each of them will lie completely within the causal future of the others.
In that case, we can obtain correlation-propagation relations relating
separately the advanced and retarded correlations to their ``propagating''
counterparts. These relations follow from the fluctuation dissipation
relations along single trajectories derived above, essentially by a method
of geometric construction : defining $\gt _{ij}^{a} = -\frac{d\mt _{ij}^{a}}
{d\t _i}$ and similarly $\gt _{ij}^{r}$, we have
\be
\gt _{ij}^{a}(\t _i,\t '_j) = \frac{e_ie_j}{4} \d (v_i(\t _i)-v_{j}(\t '_j))
\frac{dv_i}{d\t _i}.
\ee
Since the trajectory $i$ does not possess horizons, the null coordinates
$u_i$ and $v_i$ range from $-\infty $ to $\infty$. Thus the functions
$v_iv_i^{-1}$ and $u_iu_i^{-1}$ are identity functions over the entire
real line. Then we obtain, similar to equation ($5.6$),
\ba
\gt _{ij}^{a}(\t _i,\t '_j) &=& \frac{e_ie_j}{4} \d (\t _i-v_i^{-1}(v_j(\t
'_j)))
\nn \\
&=& \frac{e_j}{e_i} \gt _{ii}^{a}(\t _i, v_i^{-1}(v_j(\t '_j)))
\ea
and
\be
\gt _{ij}^{r}(\t _i,\t '_j) = \frac{e_j}{e_i} \gt _{ii}^{r}(\t _i, u_i^{-1}(u_j
(\t '_j))).
\ee
The correlations $\nt _{ij}$ may be constructed from the noises $\nt _{ii}$ in
an identical manner:
\ba
\nt _{ij}^{a}(\t _i,\t '_j) &=& \frac{e_ie_j}{4\p }\intiz \frac{dk}{k} \cos
k(v_i(\t _i)-v_i(v_i^{-1}v_j(\t '_j)))   \nn \\
&=& \frac{e_j}{e_i} \nt _{ii}^{a}(\t _i, v_i^{-1}v_j(\t '_j))
\ea
where we have inserted the identity function $v_iv_i^{-1}$ in the first step.
Also,
\be
\nt _{ij}^{r}(\t _i,\t '_j) = \frac{e_j}{e_i} \nt _{ii}^{r}(\t _i, u_i^{-1}u_j
(\t '_j)).
\ee
These two sets of constructions for the propagation and correlation kernels
in terms of the dissipation and noise kernels enables us to write down the
correlation-propagation relations simply by invoking the
fluctuation-dissipation
relations ($5.8$) as they separately apply to the advanced and retarded parts
of the
noise and dissipation along single trajectories:
\be
\nt _{ij}^{a,r}(\t _i,\t '_j) = \intin ds K_i^{a,r}(\t _i,s) \gt _{ij}^{a,r}(
s, \t '_j),
\ee
$K_i^{a}$ and $K_i^{r}$ being defined earlier ($5.9$). Since the quantities
$\gt _{ij}$ are really just $\d $-functions and the quantities $K_i^{a,r}$ are
proportional to $\nt _{ii}^{a,r}$, these relations can be
equivalently viewed as constructions of the correlations $\nt _{ij}$ from
the noises $\nt _{ii}$.

The above relations hold for trajectories without event horizons.
In the example of the uniformly accelerated detector and probe, the uniformly
accelerated detector trajectory does possess event horizons. This manifests in
the property that the range of $u_1$ is restricted to $(-\infty ,0)$ and the
range of $v_1$ to $(0,\infty )$. The probe trajectory, on the other hand, will
be
chosen to be free of horizons. We will also now assume that the probe is
switched on forever. Then we can construct the correlations $\nt _{21}$ and the
quantities
$\gt _{21}$ from $\nt _{22}$ and $\gt _{22}$ exactly as described above, and
obtain the corresponding correlation-propagation relations:
\be
\nt _{21}^{a,r}(\t _2,\t '_1) = \intin ds K_2^{a,r}(\t _2,s) \gt _{21}^{a,r}
(s, \t '_1).
\ee
This simply follows by invoking the fluctuation-dissipation relation along
the probe trajectory, as described above. However, it is of greater interest
to know whether such relations would follow from the fluctuation-dissipation
relation along the uniformly accelerated trajectory. As explained, this
will not be completely possible because the accelerated trajectory
possesses horizons. This difficulty shows up when one tries to write down
a relation of the form ($5.16$) for the quantities $\nt _{12}$ and $\gt _{12}$.
To do this, we first express the functions $Z_{ij}$ in a different form. This
was done in Section $4$ (see the steps leading from $3.54$ to $3.58$) for
the restricted case $u_2(\t _2)>0$, $v_2(\t _2)>0$. If we remove this
restriction,
we find
\ba
Z_{12}^{a} (\tau_1,\tau'_2) &=& \frac{e_1e_2}{4\p }\intiz \frac{d\o }{\o } [
\coth(\frac{\p \o }{a}) \cos \frac{\o }{a}(a\tau_1-\ln \mid av_2(\tau'_2)\mid)
\q (v_2(\t '_2))  \\
&+& \frac{\cos \frac{\o }{a}(a\t _1-\ln \mid av_2(\t '_2)\mid)}{\sinh
\frac{\p \o }{a}}\q (-v_2(\t '_2))- i\sin \frac{\o }{a} (a\t _1-\ln \mid
av_2(\t '_2)\mid)\q (v_2(\t '_2))]  \nn \\
Z_{12}^{r} (\tau_1,\tau'_2) &=& \frac{e_1e_2}{4\p }\intiz \frac{d\o }{\o } [
\coth(\frac{\p \o }{a}) \cos \frac{\o }{a}(a\tau_1+\ln \mid au_2(\tau'_2)\mid)
\q (-u_2(\t '_2))  \\
&+& \frac{\cos \frac{\o }{a}(a\t _1+\ln \mid au_2(\t '_2)\mid)}{\sinh
\frac{\p \o }{a}}\q (u_2(\t '_2))- i\sin \frac{\o }{a} (a\t _1+\ln \mid
au_2(\t '_2)\mid)\q (-u_2(\t '_2))]. \nn
\ea
$Z_{21}$ can be expressed in a similar way. From the above, we see that
the advanced (retarded) correlation for $v_2>0$ ($u_2<0$) has a thermal form,
because these correlations can be constructed simply from the noise along
the accelerated trajectory. We are therefore able to write down a correlation-
propagation relation for this part of the correlations alone. This takes the
form
\be
\q (v_2(\t '_2)) \nt _{12}^{a}(\t _1,\t '_2) + \q (-u_2(\t '_2)) \nt _{12}^{r}
(\t _1,\t '_2) = \intin ds K_1(\t _1,s) \gt _{12}(s,\t '_2)
\ee
where
\ba
\gt _{12}(\t _1,\t '_2) &=& -\frac{d\mt _{12}}{d\t _1}  \\ \nn
&=& \frac{e_1e_2}{4} [\d (\t _1-a^{-1} \ln \mid av_2(\t '_2)\mid) \q (v_2(\t '
_2))
+ \d (\t _1 + a^{-1} \ln \mid au_2(\t '_2)\mid) \q (-u_2(\t '_2))]
\ea
and
\be
K_1(\t _1,s) = \intiz \frac{dk}{\p k} \coth \pka \cos k(\t _1-s).
\ee
The single relation ($5.19$), as opposed to separate relations between the
advanced and retarded parts, is a consequence of the fact that the thermal
noise is
isotropic and therefore contains equal contributions from advanced and retarded
parts.

In the context of the analysis of section $4$, where the probe is switched
on at $u_2=0$, ($5.19$) becomes
\be
\nt _{12}^{a}(\t _1,\t '_2) = \intin ds K_1^{a}(\t _1,s) \gt _{12}^{a}(s,\t
'_2).
\ee
Viewed as a construction of $\nt _{12}^{a}$ from $\nt _{11}^{a}$, this relation
lies at the heart of the RSG cancellation in ($4.16$). Viewed alternatively
as an extension
of the thermal fluctuation-dissipation relation on the uniformly accelerated
trajectory, it thus places the role of thermal equilibrium
in the RSG cancellation on firmer ground.

We now turn to the part of the correlations which do not partake in the
correlation-propagation relation above. These are the advanced
(retarded) correlations for $v_2<0$ ($u_2>0$), containing the $\sinh^{-1}$
factors, and are not expressible in terms of the noise along the
accelerated trajectory. Rather, they represent true correlations across the
future (past) horizon. If we specialize to the case $u_2>0$ as in Section
$4$, then these are exactly the correlations which contribute to the
excitation of the probe in the guise of $(r.p.)$, equation ($4.23$). The probe
may therefore be said to be excited by free field correlations across the
future horizon.

If we specialize to the simple probe trajectory $x_2(\t _2)=0$, $t_2(\t _2)=
\t _2$, then we have $u_2(\t _2) = v_2(\t _2) = \t _2$ and the expressions
($5.18$) for $Z_{12}$ acquire a symmetric form. In this special case, we can
write down a correlation-propagation relation for the entire kernel $\nt
_{12}$,
by relating the advanced part of the correlations across the horizon to the
retarded part of the propagation kernel, and vice-versa: we then have
\be
\gt _{12}(\t _1,\t '_2) = \frac{e_1e_2}{4} [\d (\t _1-a^{-1} \ln \mid a\t
'_2\mid) \q (\t '_2)
+ \d (\t _1 + a^{-1} \ln \mid a\t '_2\mid) \q (-\t '_2)].
\ee
Choosing
\be
K'_1(\t _1, \t '_1) = \intiz \frac{dk}{\p k} [\coth \frac{\p k}{a} \cos
k(\t _1-\t '_1) + \frac{\cos k(\t _1+\t '_1)}{\sinh \frac{\p k}{a}}]
\ee
we obtain
\be
\nt _{12}(\t _1,\t '_2) = \intin ds K'_1(\t _1,s)\gt _{12}(s,\t '_2)
\ee
as a correlation-propagation relation in this special case. The above relation
cannot be geometrically
constructed from the fluctuation-dissipation relation along the single
accelerated trajectory. So far, we have not been able to show
the existence of such relations in more general cases. The extra piece
in the interpolating kernel $K'_1$ comes from correlations across the
horizon, as explained earlier.

\section{Summary and discussion}

To summarize our work and findings, we have presented  a general formalism
to treat an arbitrary number of detectors modelled as oscillators in
arbitrary kinematic states, and minimally coupled to a massless scalar
field in $1+1$ dimensions. In this approach, the scalar field has been
integrated out and the detector dynamics is described by a reduced set of
effective semiclassical stochastic equations. These equations nonetheless
contain the full quantum dynamics of the field. Our treatment can
be extended to massive fields and higher dimensions by making appropriate
changes in the two-point functions $Z_{ij}$.

We studied four examples, starting with a single inertial and uniformly
accelerated detector, mainly to illustrate the new description, and
culminating in the treatment of a uniformly
accelerated oscillator and a second oscillator which functions as a probe.
We show that there exist fluctuation-dissipation relations relating
the fluctuations of the stochastic forces on the detectors to the dissipative
forces. We discover a related set of correlation-propagation relations
between the correlations of stochastic forces on different
detectors and the retarded and advanced parts of the radiation mediated
by them.
% Finally, we point out certain unsatisfactory features of the
%minimally coupled model, and show that the derivative coupling model,
%which is obtained from the minimally coupled one by a unitary
%transformation, removes these features.

In the analysis of two inertial detectors, we find that the change in
the state of the field due to the coupling with either detector
modifies the impedance functions of both detectors, and hence their
dissipative properties. Also, this coupling introduces a mutual impedance
which describes the change in the response of one detector due to
the fluctuations of the field in the vicinity of the other one. The field
fluctuations (noise) in this case are relatively trivial, and non-trivial
effects can be ascribed mainly to the impedance functions.

In the case of the accelerated detector and the probe, on the other
hand, the noise due to field fluctuations and the field correlations
between the two trajectories play a dominant role.
Since the probe cannot causally influence the accelerated
detector, the dissipative features of this problem are relatively trivial.
Here, we find that most of the terms contributing to the response of the
probe cancel out, leaving behind a contribution that arises purely from
field correlations across the horizon. This cancellation was earlier
pointed out \cite{RSG,MPB} to be a consequence of the
identity $\ck +\ck ^{\ast}=4\ga \mid\ck \mid^2$  or variations thereof,
which is a form of fluctuation-dissipation relation. Although we utilize this
identity in our calculation, we observe, however, that this
really follows from the dissipative properties
of the accelerated detector and its free uncoupled dynamics. It
therefore does not explicitly involve the fluctuations of the field.
We point out that this cancellation can instead be understood to
follow because the correlations between the accelerated detector and
probe trajectories can be expressed partly in terms of the noise
or field fluctuations along the accelerated trajectory alone, and
also because of the isotropy of this noise. The expression of
correlation in terms of noise can be equivalently viewed as a
consequence of the correlation-propagation relations we obtain in
Section $4$, which are appropriate extensions of a generalized
fluctuation-dissipation relation directly relating field fluctuations
to dissipative properties.

A distinct feature of the influence functional formalism as used in this
paper is the assumption of an uncorrelated field-oscillator initial state.
As argued in the appendix, an uncorrelated initial state is more readily
realizable in the derivative coupling model. However, since the minimal and
derivative coupling models are dynamically equivalent, we expect our final
results to be essentially unchanged, in particular the results of detector
response in the various cases studied. The discussion of
fluctuation-dissipation
relations can be reformulated as well in a way suitable to the derivative
coupling model.

We would like to mention possible extensions of this work to other problems.
In discussions of the quantum equivalence principle \cite{QEP,Unr92}, one
compares the response of a detector moving on a geodesic trajectory in
Minkowski space, and coupled to a quantum field, to its response along a
geodesic of a spacetime with a
homogenous gravitational field. The idea is to derive a suitable
transformation on the state of the quantum field which yields the same
detector response in both cases. If one can find such a transformation,
 the equality of the detector response in both cases constitutes a test
of the validity of the quantum equivalence principle for local physical
processes. However, a homogenous gravitational field defines a global
inertial frame, and so one is inclined to believe that the equivalence
principle would hold for non-local processes as well, such as the
effective dynamics of two spatially separated detectors coupled to
the same quantum field. We plan to investigate this and related issues,
especially the implications of our findings on black hole backreaction
and information problems in later works.

\appendix

\section{Appendix}
\subsection{Infrared Problems with the Minimal Coupling Model}

The infra-red effects of the minimal coupling model are not trivial.
In fact, as we will show,
the proper treatment of this simple model, including an ultra-violet
but not an infra-red cut-off, leads us to identify a super-selection
rule which prefers a particular class of bases in the model's Hilbert
space.  Using the minimal coupling model and this preferred
class of bases is equivalent to a derivative coupling model used
by Unruh and Zurek [UZ] and
a basis of direct products of unperturbed field and oscillator states.

The MC Hamiltonian is
\be\label{MC}
H_{MC} = H_{\Phi_0} + {1\over2M}\bigl(P-\epsilon \phi(0)\bigr)^2
+{\kappa\over2}Q
 ^2\;.
\ee
It is straightforward to show that the expectation value of $H_{MC}$ for soft
photon states (i.e., low energy eigenstates of the free field Hamiltonian) has
a contribution proportional to the inverse of their unperturbed energy.
This suggests that the true low energy states of this model must have strong
correlations between the field and the oscillator.
If we reject an infra-red cut-off as  unphysical,
then we must conclude that states that are direct products of field
and oscillator states will actually have energies very much higher than that
of the true ground state of the model.

The poor behaviour of the basis of unperturbed ({\it i.e.} $\epsilon\to 0$)
energy eigenstates reflects the fact that, since there are field modes
at all frequencies, expanding around $\epsilon=0$ correctly requires degenerate
perturbation theory.
Since we are trying to set up an open quantum system, however, our choice of
basis is not merely a matter of convention.  Different bases can imply
different partitions of the complete Hilbert space into `system' and
`environment' subspaces.
In the influence functional formalism, one traces over the final states of
the environment, and assumes an initial state which is often a direct
product of system and environment.  If we change what we mean by `system' and
`environment', the final trace becomes a different operation, and
the initial state becomes a different state.

 It is not the model itself, of course, but only the naive basis that is badly
 behaved: the full Hamiltonian is quadratic, and hence equivalent to a set of
 decoupled harmonic oscillators.
 To understand the problems with the basis of unperturbed energy eigenstates,
 and to identify a better basis, we should diagonalize the full Hamiltonian.

The MC Hamiltonian may be diagonalized by defining new creation and
annihilation
operators.  If the original field and conjugate momentum operators are
$\phi(x)$
and $\pi(x)$, the diagonalizing annihilation operators are
\be  \label{ak}
a_k = {1\over\sqrt{2\hbar\pi |k|}} \left(
\int_{-\infty}^\infty\!dx\,[C(k) E_k(x) + i \sin kx] [|k| \phi(x) + i \pi(x)]
\right) + \epsilon C(k) [iQ - |k| P]\;,
\ee
where $E_k(x) = (1-{k^2M\over \kappa}) \cos kx - {\epsilon^2 k \over
2\kappa}\sin
k|x|$ and $C(k) = [ (1-{k^2M\over \kappa})^2 +
{\epsilon^4k^2\over4\kappa^2}]^{-{1\over2}}$, for a massless field.

Note that the set of field-like operators $\{a_k, a^\dagger_k\}$
diagonalizes the entire Hamiltonian.  There is no normal mode of the
coupled system which corresponds even weakly to the unperturbed
oscillator, and all of the normal modes contain very non-local
excitations of the field.  If we wish to consider the oscillator as an
open system coupled to the field as an unobserved environment, then,
the basis of exact energy eigenstates of the combined system will not
be particularly convenient.  The fact that this model is easily solved
exactly does not make the system-environment problem trivial.

We can now, however, determine the effect on the true ground state of
the unperturbed oscillator raising and lowering operators.  We find
that either of these operators maps the ground state onto a highly
excited state, whose expected energy is infra-red divergent.
Consequently, observations that are restricted to the oscillator sector
alone (as it is defined from Equation \(MC) may be said to require
infinite amounts of energy.  Considering the field that appears in
Equation \(MC) to be an unoberved environment is therefore unphysical.

We can, however, change our basis so that the Hamiltonian appears more
benign in terms of the transformed operators.  We may effect this
transformation using the unitary operator $U= \exp
-{i\over\hbar}\epsilon Q \phi(0)$.  This transformation mixes the field
and oscillator sectors, and so changes what is meant by an observation
of the oscillator alone.  We can check that the new oscillator raising
and lowering operators, acting on the ground state, now produce states
whose energy is ultraviolet divergent, instead of infra-red.  A
physically plausible UV cut-off then renders this energy finite, and it
becomes reasonable to consider the new field sector as unobserved.
Furthermore, a direct product of the unperturbed ground state of the
transformed field and any finite-energy oscillator state now has finite
energy, and so is not unreasonable as an initial state for the coupled
system.

An alternative way of expressing the advantage of the transformed
variables is to say that in order for a degree of freedom in the theory
to be observable in isolation, it must require finite energy to excite
or de-excite it without affecting other degrees of freedom, and it must
also be spatially local (except possibly at small scales).  Such an
observed degree of freedom will be some linear combination of the exact
normal modes.  If we require the co-efficients in this linear
combination to vanish at high energies and remain finite everwhere
else, we can ensure finite energy observability.  If we require that
the co-efficients are constant at low energies, we also ensure local
observability.

We would then like to have a basis in which this observed linear
combination ``looks like'' a harmonic oscillator coupled to a scalar
field.  Given our conditions of local and finite energy observability,
our original basis does not provide this feature, but our second basis
does.

The transformed Hamiltonian $H_{MC}$ becomes, in the new operators,
precisely the Hamiltonian of the Unruh-Zurek model.  We have therefore
found that, even if we begin by analysing the minimally-coupled model,
we may be compelled in the end to study the Unruh-Zurek model (with a UV
cut-off) instead.  More convenient for our subsequent calculations than
the Hamiltonian for this model is its Lagrangian:
\be  \label{Luz}
L_{DC} = {1\over2}\left(M\dot Q^2 - \kappa Q^2\right) + {1\over2} \int\!dk\,
\dot\Phi_k^*\dot\Phi_k - k^2 \Phi_k^*\Phi_k - {\epsilon\over\sqrt{2\pi}}Q
(\dot \Phi_k + \dot\Phi_k^*)\;.
\ee
Here $\Phi_k$ is the time-dependent, spatial Fourier transform of the field
$\phi$.

Note that, even in the more benign Unruh-Zurek basis, distinguishing
the oscillator as a system observable independently of the field
directly implies that there must be a UV cut-off.  One often argues
that a cut-off is appropriate because one is not interested in
accurately describing physics at inaccessible energy scales; but in the
case of the oscillator coupled to a field, there must really be a
cut-off in the coupling in order for there to be any accessible energy
scales!

\subsection{Correspondence between MC and derivative coupling models}

In the minimal coupling model, the derivative of the oscillator
coordinate couples to the field, whereas in the derivative coupling
model the oscillator coordinate couples to the derivative of the
field. These two models thus differ by a total derivative term in the
Lagrangian. In particular, they have the same Heisenberg operator
dynamics. The above subsection describes the issue of the initial
state, and argues that an uncorrelated initial state is physically
more realistic in the derivative coupling model. However, since the
two models have the same dynamics, this should translate to a simple
prescription for switching from one model to the other in the
context of the influence functional treatment.

In the previous sections, we have derived all results from the
minimal coupling model. One can obtain corresponding quantities in the
influence functional of the
derivative coupling model via the prescription below:
\ba
\frac{dQ_i}{d\t _i}&\rightarrow& Q_i  \nn \\
\frac{dQ'_i}{d\t _i}&\rightarrow& Q'_i  \nn \\
Z_{ij}(\t _i,\t '_j)&\rightarrow& \frac{d}{d\t _i} \frac{d}{d\t '_j}Z_{ij}(
\t _i,\t '_j)
\ea
The stochastic effective action in the derivative coupling model is then given
by
\ba
S_{eff} &=& \sum_{i=1}^{N} \inti d\t _i [ \dot{Q^-_i} \dot{Q^+_i} -
\O _{i}^{2} Q^-_iQ^+_i
- Q^-_i s_i(\t _i)\eta_i(\t _i) \nn \\
& &- 2 Q^- _i
s_i(\t _i) \sum_{j=1}^{N} \intij d\t '_j Q^+_{j'} s_j(\t '_j)
\mt _{ij}(\t _i,\t '_j) ]    .
\ea
Note that the quantities $\mt _{ij}$ in the above equation refer to the
newly defined quantities in the derivative coupling model. They are obtained
by differentiating the corresponding quantities in the MC model twice.

The Langevin equations are:
\be
\frac{d^2 Q_i}{d\t _i^{2}} - 2\sum_{j=1}^{N} \intij d\t '_j s_j(\t '_j)
 s_i(\t _i)\mt _{ij}(\t _i,\t '_j)  Q_j
+ \O _i^2 Q_i = s_i(\t _i) \eta_i(\t _i).
\ee
The noise kernel, as the correlator of $\eta_i$ and $\eta_j$, is also
obtained by the corresponding noise kernel in the MC model by
differentiating twice, according to the correspondence established
above.

The infrared divergent energy of the initially uncorrelated state does
have an effect: the propagation kernel, in the MC model, contains an
initial ``shock wave'' term, as well as the expected dissipation and
propagation terms, and this term is not present in the DC model.
Since this shock wave is a transient, it has no
significance in our late-time analysis.\\

\noindent {\bf Acknowledgement}
This work is supported in part by the National Science Foundation
under grants PHY91-19726.
AR would like to thank Professor Ulrich Gerlach for
the opportunity to present an early version of this work at the Seventh
Marcel Grossmann Meeting in General Relativity, 1994 and the University of
Maryland Graduate School for partial travel support.
BLH acknowledges support from the General Research Board of the
Graduate School of the University of Maryland and the Dyson Visiting Professor
Fund at the  Institute for Advanced Study, Princeton.
Part of this work was done while he visited the
Newton Institute for Mathematical Sciences at the University of Cambridge
during the Geometry and Gravity program in Spring 1994.
% and the physics department of the Hong Kong University of Science and
% Technology in Spring 1995.
JA thanks Salman Habib for discussions, and the Canadian Natural Sciences and
Engineering Research Council for support.

\end{document}